\documentclass[prd,aps,showpacs,nofootinbib]{revtex4}
\usepackage{amssymb}
\usepackage{graphicx}
\usepackage{amsmath}

\newcommand{\be}{\begin{equation}}
\newcommand{\ee}{\end{equation}}
\newcommand{\bq}{\begin{eqnarray}}
\newcommand{\eq}{\end{eqnarray}}

\begin{document}
\title{Lorentz violation with an invariant minimum speed as foundation of the tachyonic inflation within a Machian scenario} 
\author{**Cl\'audio Nassif Cruz, *Rodrigo Francisco dos Santos and A. C. Amaro de Faria Jr.}

\affiliation{\small{{\bf **CPFT}: {\bf Centro de Pesquisas em F\'isica Te\'orica}, Rua Rio de Janeiro 1186, Lourdes, CEP:30.160-041,
 Belo Horizonte-MG, Brazil.\\
 {\bf *UFF: Universidade Federal Fluminense}, Av.Litoranea s/n, Gragoat\'a, CEP:24210-340, Niter\'oi-RJ, Brazil.\\
 {\bf IEAv: Instituto de Estudos Avan\c{c}ados}, Rodovia dos Tamoios Km 099, 12220-000, S\~ao Jos\'e dos Campos-SP, Brazil.\\
 **claudionassif@yahoo.com.br, *santosst1@gmail.com, antoniocarlos@ieav.cta.br}} 

\begin{abstract}
We show the relationship between the scalar kinematics potential of Symmetrical Special Relativity (SSR) and the ultra-referential of
vacuum connected to an invariant minimum speed postulated by SSR. The property of the conformal metric of SSR is showed, from where we 
deduce a kind of de Sitter metric. The negative curvature of spacetime is calculated from the conformal property of the metric. Einstein 
equation provides an energy-momentum tensor which is proportional to SSR-metric. We also realize that SSR leads to a deformed kinematics 
with quantum aspects directly related to the delocalization of the particle and thus
being connected to the uncertainty principle. We finish this work by identifying the lagrangian of SSR with the so-called tachyonic models
(slow-roll), where the tachyonic potential is a function depending on the conformal factor, thus allowing the SSR-lagrangian to be able 
to mimic a tachyonic lagrangian related to the so-called Dirac-Born-Infeld lagrangian, where the superluminal effects are interpreted 
as being a large stretching of spacetime due to new relativistic effects close to the invariant minimum speed as being the foundation 
of the inflationary vacuum connected to a variable cosmological parameter that recovers the cosmological constant for the current 
universe.  
\end{abstract}


\maketitle

\section{Introduction}

In many recent experiments, one has studied the working of the so-called acoustic black holes. These acoustic black holes would be
fluid-flowing holes (colloids)\cite{acus, acus1}. The interesting thing about studying this type of system as a toy model for a quantum
gravitation is precisely to be able to study alternative forms of causal structures (sound cones), which may represent new causal 
structures in scenarios where the vacuum is represented by a colloid, so that the vacuum has some dynamics. These studies abandon the idea of a
static vacuum. Thus the vacuum is seen as a cosmological fluid, which is similar to a colloid \cite{acus2}. Ideal gas models are widely 
investigated in cosmology. Fifth-essence, phantom fluids, Chaplegin's gas\cite{li}\cite{Chap}, inflationary fluids\cite{Guth, Volo} are 
generated by several theories\cite{Sergio, Sergio1, Sergio2}, without these various fluids being unified in a single theoretical framework. 
Usually models of repulsive fluids are related to de Sitter relativity\cite{Aldro1, Aldro2, Aldro3}. However, they are not still 
unified into a unique scenario.

SSR-theory has the potential to unify all of these fluids, such as the gravito-electrical coupling regimes governed by the new 
dimensionless constant of nature $\xi$\cite{N2016}, working like a ultra-fine structure constant of gravito-electromagnetic origin. Until 
the present date, SSR has been able to generate the equation of state (EOS) of vacuum, i.e.,  
$p=-\rho$\cite{N2016, N2015, N2012, N2010}. However, the gravitational phase transition between the repulsive and attractive gravity,
which occurs at $v=v_0=\sqrt{cV}$ in SSR, has not been deeply discussed yet. This work intends precisely to describe firstly the 
slightly repulsive regime of gravity associated with the constant $\xi$, where Lorentz symmetry can be easyly recoverd. After, the regime of
stronger repulsion associated with the constant $\xi^2$ is considered\cite{N2016}. We can still go further by exploring a chaotic inflation 
regime, where Lorentz symmetry is completely destroyed. This scheme has been already showed\cite{Rodrigo} when we calculate the curvature
of the space generated by the SSR and we verify that such a space has not a constant curvature. The variation of the curvature is
explained, because it is possible to assemble a SSR-lagrangian, which imitates a slow roll (tachyon)\cite{li}\cite{Guth} structure,
associated with its kinematic scalar field. Therefore the potential $\phi$ of SSR generates a tachyonic potential, being associated with the cosmic inflation. 
This construction shows that SSR has a rich structure not fully explored yet.

\section{A brief review of the causal structure of SSR}

\begin{figure}
\begin{center}
\includegraphics[scale=0.90]{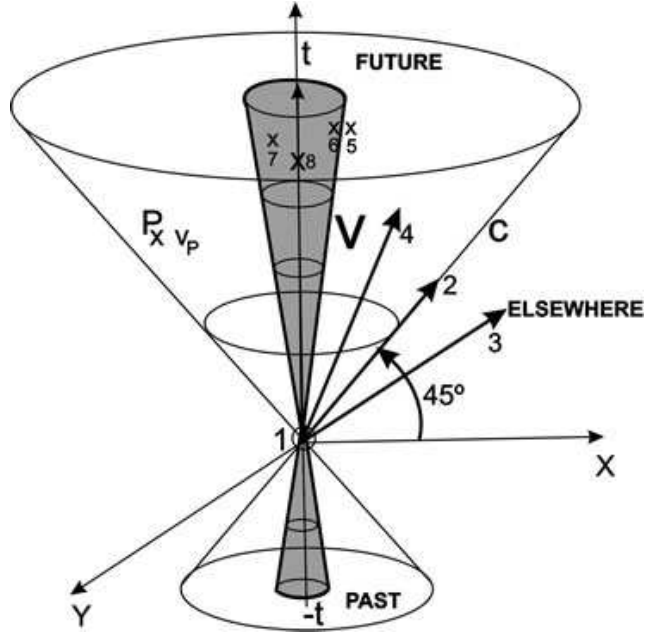}
\end{center}
\caption{The external and internal conical surfaces represent respectively the speed of light $c$ and the unattainable minimum
speed $V$, which is a definitely prohibited boundary for any particle. For a point $P$ in the world line of a particle, in the interior
of the two conical surfaces, we obtain a corresponding internal conical surface, such that we must have $V<v_p\leq c$. The $4$-interval
$S_4$ is a time-like interval. The $4$-interval $S_2$ is a light-like interval (surface of the light cone). 
The $4$-interval $S_3$ is a space-like interval (elsewhere). The novelty in spacetime of SSR are the $4$-intervals $S_5$ (surface of
the dark cone) representing an infinitly dilated time-like interval, including the $4$-intervals $S_6$, $S_7$ and $S_8$ inside the dark 
cone for representing a new space-like region (see ref.\cite{N2016}).}
\end{figure}

The breakdown of Lorentz symmetry for very low energies\cite{N2016}\cite{N2015} generated by the presence of a
background field is due to an invariant mimimum speed $V$ and also a universal dimensionless constant $\xi$\cite{N2016}, working
like a gravito-electromagnetic fine constant, namely:

\begin{equation}
\xi=\frac{V}{c}=\sqrt{\frac{Gm_{p}m_{e}}{4\pi\epsilon_0}}\frac{q_{e}}{\hbar c},
\end{equation}
$V$ being the minimum speed and $m_{p}$ and $m_{e}$ are respectively the mass of the proton and electron. Such a minimum 
speed is $V=4.5876\times 10^{-14}$ m/s. We have found $\xi=1.5302\times 10^{-22}$, where Dirac's large number hypothesis (LNH) were
taken into account in obtaining $\xi$\cite{N2016}. 

It was shown\cite{N2016} that the minimum speed is connected to the cosmological constant in the following way:

\begin{equation}
V\approx\sqrt{\frac{e^{2}}{m_{p}}\Lambda^{\frac{1}{2}}},
\end{equation}
where $\Lambda=6c^2/r_H^2\sim 10^{-35}s^{-2}$\cite{N2016}, whose tiny value is in good agreement with the observational data. $r_H\sim 10^{26}$m is the Hubble radius. Thus SSR was 
able to provide the tiny positive value of the current cosmological constant, where the quantum field theories 
(QFTs) fail, since QFTs predict a discrepance of about $120$ orders of magnitude for the cosmological constant. This is  
well-known as the {\it Cosmological Constant Puzzle} treated by SSR\cite{N2016}\cite{N2015}. 

Therefore we realize that the light cone contains a new region of causality called {\it dark cone}\cite{N2016}, so that the speed of a
particle must belong to the following range: $V$(dark cone)$<v<c$ (light cone) (Fig.1). 

The breaking of Lorentz symmetry group destroys the properties of the transformations of Special Relativity (SR) and so generates
an intriguing kinematics and dynamics for speeds very close to the minimum speed $V$, i.e., for $v\rightarrow V$, we find new 
relativistic effects such as the contraction of the improper time and the dilation of space\cite{N2016}. In this new scenario,
the proper time also suffers relativistic effects such as its own dilation with respect to the improper one, i.e., $\Delta\tau>>\Delta t$ 
when $v\rightarrow V$, so that we have found

\begin{figure}
\includegraphics[scale=0.90]{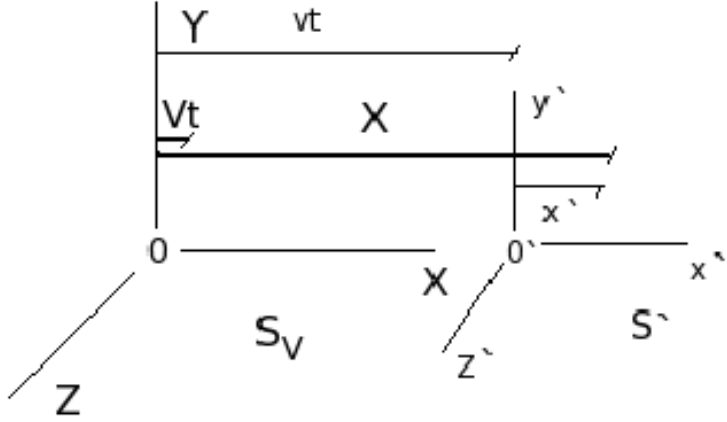}
\caption{In this special case of $(1+1)D$, the referential $S^{\prime}$ moves in $x$-direction with a speed $v(>V)$ with respect to the
 background field connected to the ultra-referential $S_V$. If $V\rightarrow 0$, $S_V$ is eliminated (empty space), and thus the galilean
 frame $S$ takes place, recovering Lorentz transformations.}
\end{figure}

\begin{equation}
\Delta\tau\sqrt{1-\frac{V^{2}}{v^{2}}}=\Delta t\sqrt{1-\frac{v^{2}}{c^{2}}},
\end{equation}
which was shown in the reference\cite{N2016}, where it was also made experimental prospects\cite{N2018} for detecting such new relativistic 
effect close to the invariant minimum speed $V$, i.e., too close to the absolute zero temperature. 

Since the minimum speed $V$ is an invariant quantity as the speed of light $c$, $V$ does not alter the value of the speed $v$ of
any particle. Therefore we have called ultra-referential $S_{V}$\cite{N2016}\cite{N2015} as being the preferred (background) reference
frame in relation to which we have the speeds $v$ of any particle. In view of this, the reference frame transformations change 
substantially in the presence of the preferred frame $S_V$, as follows: 

a) The special case of $(1+1)D$ transformations in SSR\cite{N2016}\cite{N2015}\cite{N2012}\cite{N2010} 
with $\vec v=v_x=v$ (Fig.2) are 

\begin{equation}
x^{\prime}=\Psi(X-vt+Vt)=\theta\gamma(X-vt+Vt) 
\end{equation}

and 

\begin{equation}
t^{\prime}=\Psi\left(t-\frac{vX}{c^2}+\frac{VX}{c^2}\right)=\theta\gamma\left(t-\frac{vX}{c^2}+\frac{VX}{c^2}\right), 
\end{equation}
where $\theta=\sqrt{1-V^2/v^2}$ and $\Psi=\theta\gamma=\sqrt{1-V^2/v^2}/\sqrt{1-v^2/c^2}$.

b) The $(3+1)D$ transformations in SSR (Fig.3)\cite{N2016} are

\begin{equation}
\vec{r'}=\theta\left[\vec{r_{T}}+\gamma\left(\vec{r_{//}}-\vec{v}\left(1-\frac{V}{v}\right)t\right)\right]=
\theta\left[\vec{r_{T}}+\gamma\left(\vec{r_{//}}-\vec{v}t+\vec{V}t\right)\right]
\end{equation}

and

\begin{equation}
t'=\theta\gamma\left[t-\frac{\vec{r}\cdotp\vec{v}}{c^{2}}+\frac{\vec{r}\cdotp\vec{V}}{c^{2}}\right]. 
\end{equation}

Of course, if we make $V\rightarrow 0$, we recover the well-known Lorentz transformations.

\begin{figure}
\includegraphics[scale=0.90]{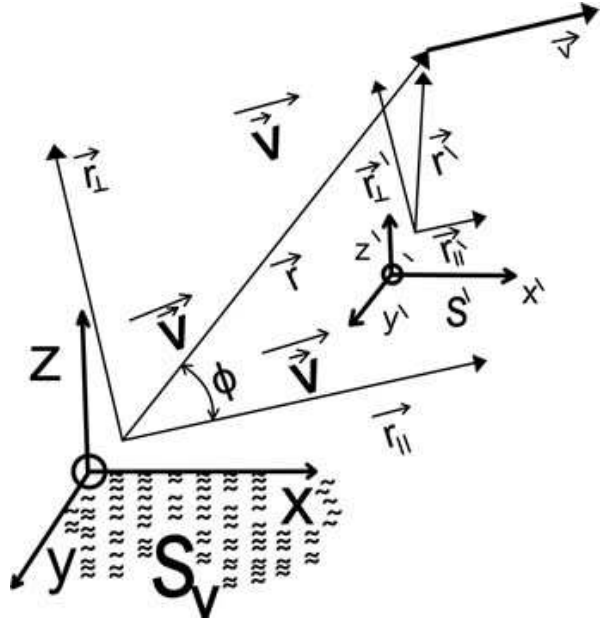}
\caption{$S^{\prime}$ moves with a $3D$-velocity $\vec v=(v_x,v_y,v_z)$ in relation to $S_V$. For the special case of $1D$-velocity
$\vec v=(v_x)$, we recover Fig.2; however, in this general case of $3D$-velocity $\vec v$, there must be a background vector $\vec V$
(minimum velocity) with the same direction of $\vec v$ as shown in this figure. Such a background vector $\vec V=(V/v)\vec v$ is 
related to the background reference frame (ultra-referential) $S_V$, thus leading to Lorentz violation. The modulus of $\vec V$ is 
invariant at any direction.} 
\end{figure}

Although we associate the minimum speed $V$ with the ultra-referential $S_{V}$, this frame is inaccessible for any particle. Thus, the
effect of such new causal structure of spacetime generates an effect on mass-energy being symmetrical to what happens close to the speed 
of light $c$, i.e., it was shown that $E=m_0c^2\Psi(v)=m_0c^2\sqrt{1-V^2/v^2}/\sqrt{1-v^2/c^2}$, so that $E\rightarrow 0$ when
 $v\rightarrow V$\cite{N2016}\cite{N2015}\cite{N2012}. We notice that $E=E_0=m_0c^2$ for $v=v_0=\sqrt{cV}$. In a previous 
paper\cite{N2016}, it was also shown that the minimum speed $V$ is associated with the cosmological constant, which is equivalent to a
fluid (vacuum energy) with negative pressure so that its EOS is $p=-\rho$\cite{N2016}\cite{N2015}. This EOS naturally emerged 
from first component $T^{00}$ of the energy-momentum tensor of perfect fluid of SSR when the limit $v\rightarrow V$ was taken 
into account on $T^{\mu\nu}$ without {\it ad-hoc} conditions, i.e., we just have considered the tensor of perfect fluid 
$T^{\mu\nu}=(p+\rho)\mathcal U^{\mu}\mathcal U^{\nu}-pg^{\mu\nu}$, where $\mathcal U^{\mu}=\Psi(v)[1,v_{\alpha}/c]$ with $\alpha=1,2,3$
is the contravariant $4$-velocity of SSR. So by calculating the first component $T^{00}$ by using such $4$-velocity of SSR, we have 
found $T^{00}=[\rho(1-V^2/v^2)+p(v^2/c^2-V^2/v^2)]/(1-v^2/c^2)$\cite{N2016}\cite{N2015}, such that, in the vacuum regime, 
we have obtained $\lim_{v\rightarrow V} T^{00}=\rho_{vac}=p(\xi^2-1)/(1-\xi^2)=-p$, where $\xi=V/c$, or then $p=-\rho_{vac}$, 
which is exactly the EOS for the vacuum energy, which represents the negative pressure for the vacuum energy density related to
the cosmological constant. Therefore we have shown that there is a relationship between the invariant minimum speed, the vacuum
energy density and the cosmological constant whose positive tiny value was obtained by SSR, being in agreement with 
the observational data\cite{N2016}\cite{N2015}\cite{N2008}.     

The metric of such symmetrical spacetime of SSR is a deformed Minkowski metric with a global multiplicative function (a scale factor
with $v$-dependence) $\Theta(v)$ working like a conformal factor, which leads us to an extended dS-metric\cite{Rodrigo}. Thus we write 

\begin{equation}
ds^{2}=\Theta\eta_{\mu\nu}dx^{\mu}dx^{\nu},
\end{equation}
where $\Theta=\Theta(v)=\theta^{-2}=1/(1-V^2/v^2)\equiv 1/(1-\Lambda r^2/6c^2)^2$\cite{Rodrigo}, $\Lambda$ being the cosmological 
parameter and $\eta_{\mu\nu}$ being the Minkowski metric.

We can say that SSR geometrizes the quantum phenomena as investigated before (the origin of the Uncertainty Principle)\cite{N2012} 
in order to allow us to associate quantities belonging to the microscopic world with a new geometric structure that originates from the
Lorentz symmetry breaking due to an invariant minimum speed.  

\section{The scalar field and the metric of SSR}

 Let's write the energy of a particle in SSR in the following way: 
 
\begin{equation}
 E=m_{0}c^{2}\frac{dt}{d\tau}, 
 \end{equation}

where we have 
 
\begin{equation}
 \Psi=\Psi(v)=\frac{dt}{d\tau}=\frac{\sqrt{1-\frac{V^2}{v^2}}}{\sqrt{1-\frac{v^2}{c^2}}}, 
\end{equation}
so that the energy $E$ by taking into account the ultra-referential $S_{V}$ is simply written as 
  
\begin{equation}
 E=m_{0}c^{2}\Psi
\end{equation}
We can now define the potential of SSR, as follows: 
\begin{equation}
 \phi=c^{2}\left[\Psi(v)-1\right], 
\end{equation}
such that we find $\Psi(v)\equiv(1+\phi/c^2)$, where $-c^2\leq\phi<\infty$. 
 
By following Nassif\cite{N2016}, let us show the scalar field connected to the vacuum represented by $S_{V}$. We should 
realize that such a field is a strongly repulsive scalar pontential, so that $\phi<<0$. This is a non-classical aspect of 
gravity for much lower energies given in the limit of vacuum, i.e., $v\approx V$. Thus we write the following approximation
by neglecting the Lorentz factor (higher energies), namely: 

\begin{equation}
 E\approx{m_{0}}c^{2}\Theta^{-\frac{1}{2}}\approx{m_{0}}c^{2}\left(1+\frac{\phi}{c^2}\right),
\end{equation}
where $\Theta^{-1/2}=\theta=\sqrt{1-V^2/v^2}$. In the limit $v\rightarrow V$, this implies $E\rightarrow 0$, which
corresponds to $\phi\rightarrow-c^2$, i.e., this is the lowest potential for representing the most repulsive potential related to the
vacuum $S_V$. Thus we find $\phi(V)=-c^2$ for representing the most fundamental vacuum potential. We have interpreted this result\cite{N2016} for
describing an exotic particle (an infinitely massive boson) of vacuum that would escape from a maximum anti-gravity $\phi=-c^2$, but 
any finite massive particle could escape from such strongest anti-gravity working like an anti-gravitational horizon 
connected to a de Sitter horizon associated with a sphere of dark energy\cite{Rodrigo}. So it is interesting to note that the virtual
particle of vacumm in $S_V$ has infinite mass since $\Theta^{-1}(v=V)=0$, thus being the counterpart of the photon, which is a massless 
particle as $\gamma(v=c)=\infty$. 

Nassif\cite{N2016} has demonstrated the relationship between the most negative gravitational potential in SSR ($\phi$) with the
cosmological constant, namely: 

\begin{equation}
 \phi_{\Lambda}=\phi(V)=-\frac{\Lambda r^2}{6}=-c^2, 
\end{equation}
from where it was obtained a cosmological parameter as being $\Lambda(r)=6c^2/r^2$, with $\Lambda>0$, $r$ being the radius of a sphere
of dark energy. 

We have realized that the scalar field is directly related to the conformal factor, as follows: 

\begin{equation}
 \Theta^{-1}(v)=\left(1-\frac{V^2}{v^2}\right)\equiv\Theta^{-1}(\phi)=\left(1+\frac{\phi}{c^2}\right)^2,
\end{equation}
where we have $\phi=-\Lambda r^2/6$\cite{Rodrigo}.

Therefore, we can write the scalar potential $\phi$ just for the repulsive sector $V<v<v_{0}(=\sqrt{cV})$ as being the following
approximation: 

\begin{equation}\label{tac}
 \phi(v)=-c^{2}\left(1-\sqrt{1-\frac{V^2}{v^2}}\right), 
\end{equation}
where we have neglected the Lorentz factor $\gamma$. 

As we can see above, the kinematics structure of the spacetime in SSR is the generator of the repulsive gravitational potential 
that plays the role of a positive cosmological parameter $\Lambda(r)$ that recovers the current cosmological constant for 
$r=r_H\sim 10^{26}m$, i.e., 
$\Lambda_0=\Lambda(r_H)=6c^2/r_H^2\sim 10^{-35}s^{-2}$\cite{N2016}. 

 When we just take into account a weak gravitational repulsion, we get $v\approx v_{0}=\sqrt{Vc}$, so that we write

\begin{equation}\label{escalar}
 \phi(v\approx{v_{0}}=\sqrt{Vc})=-c^{2}\left(1-\sqrt{1-\frac{V}{c}}\right)\approx 0^{-},
\end{equation}
which represents a limit of very weak repulsion close to the phase transition between repulsive and attractive gravity. 

\begin{figure}     
\centering
\includegraphics[scale=0.90]{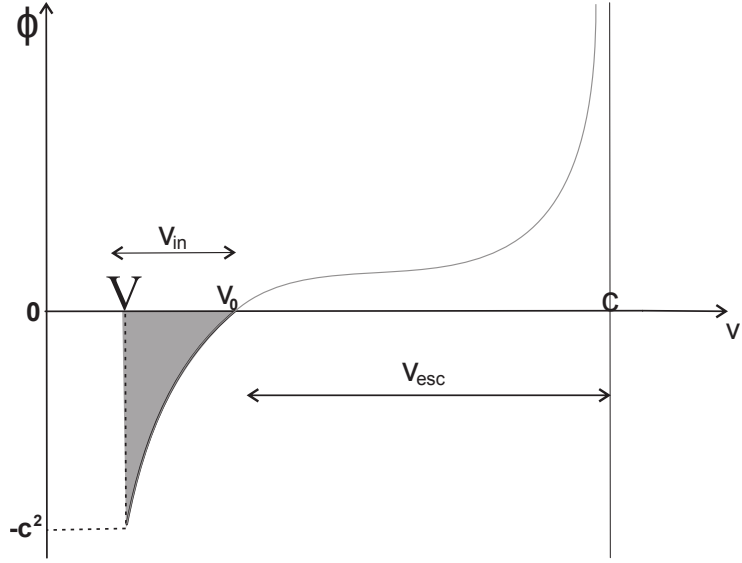}
\caption{Graph showing the general scalar potential $\phi(v)=c^{2}\left(\sqrt{\frac{1-\frac{V^2}{v^2}}{1-\frac{v^2}{c^2}}}-1\right)$
 given in function of the velocity. It shows the two phases, namely gravity (right side)/anti-gravity (left site), where the barrier
at the right side represents the ultra-relativistic limit (speed of light $c$ with $\phi\rightarrow\infty$), and the barrier at the 
left side represents the quantum (anti-gravitational) limit only described by SSR (minimum speed $V$ with $\phi=-c^2$). The long
intermediary region is the Newtonian regime ($V<<v<<c$), where occurs the phase transition just for $v=v_0=\sqrt{cV}$.}
\label{Rotulo}
\end{figure}

Let us now consider a conformal transformation $\omega^2$\cite{Rodrigo}, which is a scale transformation on a metric given in the metric space, thus
leading to a new metric connected to another metric space, namely $\mathcal{G}_{\mu\nu}=\omega^{2}{g}_{\mu\nu}$, which must obey the 
following conditions: 
 
i) $\omega^2$ presents inverse, i.e., $\omega^{-2}$ being a soft function, which generetes all its derivative functions. 
It transforms the neighborhoods of a point $P$, in the topological sense, leading to the neighborhood of a point $P^{\prime}$ 
in the transformed metric space. This is equivalent to the isometry properties.

ii) The transformation $\omega^2$ preserves the zero geodesic, namely: 

\begin{equation}
 g_{\mu\nu}X^{\mu}X^{\nu}=g_{\mu\nu}(\omega{x^{\mu}})(\omega{x^{\nu}})=\omega^{2}g_{\mu\nu}x^{\mu}x^{\nu}=
\mathcal{G}_{\mu\nu}x^{\mu}x^{\nu}=0, 
\end{equation}
which means that the transformation $X^{\mu}=\omega{x^{\mu}}$ preserves the space, time and light-like vectors, which also
implies that $\omega>0$. 

iii) It must obey the conservation of angles, namely:

\begin{equation}
 \frac{1}{\sqrt{|w||u|}}g_{\mu\nu}w^{\mu}u^{\nu}=\frac{1}{\sqrt{|W||U|}}\mathcal{G}_{\mu\nu}W^{\mu}U^{\nu}, 
\end{equation}
where $\vec w$, $\vec u$, $\vec W$ and $\vec U$ are any vectors. 

iv) We should consider $\omega(a)$ such that the parameter $a$ needs to belong to the original metric space. 

In order to verify the conditions above on the scale transformation of SSR, where $\mathcal{G}_{\mu\nu}=\Theta(v)\eta_{\mu\nu}$, we
see that $\Theta(v)=\frac{1}{\left(1-\frac{V^2}{v^2}\right)}$ has inverse, so that we find $\Theta^{-1}=\left(1-\frac{V^2}{v^2}\right)$.
This function is defined in the interval $V<v<c$ such that $0<\Theta^{-1}<1-\xi^{2}$, its inverse being in the interval
$\frac{1}{1-\xi^{2}}<\Theta(v)<\infty$, where the speed $v$ is given with respect to the ultra-referential $S_V$\cite{N2016}. 

In order to realize that $v$ is defined in the metric space of SSR, we should have in mind that $v$ is directly related to measurable 
greatnesses like energy given in the dispersion relation, as follows: 

\begin{equation}
 \Theta\eta_{\mu\nu}\mathcal P^{\mu}\mathcal P^{\nu}=m_{0}^{2}c^{2},  
\end{equation}
from where we get

\begin{equation}
 \mathcal P_{\mu}\mathcal P^{\mu}=m_{0}^{2}c^{2}\left(1-\frac{V^2}{v^2}\right),  
\end{equation}
where $\mathcal P^{\mu}=m_0c\mathcal U^{\mu}$ and  $\mathcal P_{\mu}=m_0c\mathcal U_{\mu}$, with 
$\mathcal U^{\mu}=\Psi(v)[1,v_{\alpha}/c]$ and $\mathcal U_{\mu}=\Psi(v)[1,-v_{\alpha}/c]$. 

Now, verifying the null geodesic, i.e., 

\begin{equation}
 \eta_{\mu\nu}x^{\mu}x^{\nu}=0, 
\end{equation}
and having $\theta=\sqrt{\left(1-\frac{V^2}{v^2}\right)}$, so we find $X^{\mu}=\sqrt{\left(1-\frac{V^2}{v^2}\right)}x^{\mu}$. Thus, by
applying such transformation to the null geodesic, we obtain 

\begin{equation}
 \eta_{\mu\nu}\sqrt{\left(1-\frac{V^2}{v^2}\right)}x^{\mu}\sqrt{\left(1-\frac{V^2}{v^2}\right)}x^{\nu}=
\left(1-\frac{V^2}{v^2}\right)\eta_{\mu\nu}x^{\mu}x^{\nu}=0. 
\end{equation}

Of course the angle conservation is obeyed by previous construction. In SR we have the well-known {\it boosts}, but in SSR, the 
``boosts'' in the vicinity of the minimum speed ($v{\approx}V$) are deformed in the following way: Consider the symmetric matrix
${\theta}I$\cite{N2015}\cite{N2016}, so that a vector in the interval $ds$ becomes invariant under the scale transformation, as follows: 

\begin{equation} 
 x^{*\mu}={\theta}Ix^{\mu}=\sqrt{\left(1-\frac{V^2}{v^2}\right)}Ix^{\mu}, 
\end{equation}
 where $\theta=\Theta^{-1/2}$ is also a scale factor. Alternatively we have written 
$\Lambda(v{\approx}V)x^{\mu}=x^{*\mu}$\cite{N2015}\cite{N2016}. 

 We write

  \begin{equation}
   (ds^{*})^{2}=ds^{2}(v)=dx^{*\mu}dx^{*}_{\mu}=\theta^2ds^{2}, 
  \end{equation}
where $ds^{2}$ is the squared interval of the usual space-time in SR and $(ds^{*})^{2}$ is the deformed interval that dilates
drastically close to the minimum speed. Thus the effect of the introduction of a minimum speed with the presence of the
ultra-referential $S_{V}$ leads to a scale deformation, so that the deformed interval $(ds^{*})^{2}$ remains invariant. In view of
this, we write 

\begin{equation}
 (ds'*)^{2}=(ds*)^{2}=\eta_{\mu\nu}dx*^{\mu}dx*^{\nu}, 
\end{equation}
where $(ds*)^{2}=(1-\alpha^2)dx^{\mu}dx_{\mu}=\theta^{2}ds^{2}$. Of course in the limit of $V\rightarrow 0$ ($\alpha\rightarrow0$), 
the effect of scale transformation vanishes and so SSR recovers SR-theory. 

Now we write SSR-metric, as follows: 

\begin{equation}
 ds^{2}={\Theta}\eta_{\mu\nu}dx^{\mu}dx^{\nu}, 
\end{equation}
where

\begin{equation}
 \Theta=\Theta(v)=\frac{1}{\left(1-\frac{V^2}{v^2}\right)}. 
\end{equation}

Thus the covariant SSR-metric is under the following scale transformation:  

\begin{equation}
 \mathcal G_{\mu\nu}=\Theta\eta_{\mu\nu}=\frac{1}{\left(1-\frac{V^2}{v^2}\right)}\left[\begin{array}{rrcccccccc}
 1 & 0    & 0  & 0 \\
 0   & -1 & 0    & 0 \\
 0   & 0 & -1 & 0 \\
 0   & 0 & 0 & -1 \\
\end{array} \right]. 
\end{equation}

It is easy to verify that we recover Minkowski metric ($\eta_{\mu\nu}$) in the limit $V\rightarrow 0$. Thus we have realized that the
scale transformation of SSR ($\Theta$) generates a conformally flat metric, i.e., $\Theta\eta_{\mu\nu}$\cite{Rodrigo}.  

We have $\phi=-\frac{\Lambda{r}^2}{6}$\cite{N2016} in order to obtain 

\begin{equation}\label{metsitter}
d\mathcal{S}^{2}=\frac{c^{2}dt^2}{\left(1-\frac{\Lambda r^2}{6c^2}\right)^2}-\Sigma_{i=1}^{3}\frac{dx^{2}_{i}}
{\left(1-\frac{\Lambda r^2}{6c^2}\right)^2}
\end{equation}

 We see that the metric in Eq.(30) is very similar to de Sitter metric\cite{Rodrigo}. 
 
\section{Negative curvature}

 Knowing the negative curvature of the spacetime of SSR, we can calculate the Ricci tensor. It is expected that the cosmological constant 
introduces a negative curvature in spacetime\cite{tes}. Thus a de Sitter geometry should emerge from our model\cite{Rodrigo}. Here SSR has 
a conformal structure with negative curvature. Such negative curvature should be more negative when the speed is decreased even more,  
leading to a delocalization of the particle. So we realize that the increase of delocalization (uncertainty) on position of the 
particle is directly connected to the increase of the modulus of the negative curvature, showing that a stronger anti-gravitational 
potential close to $-c^2$ ($v\rightarrow V$) would be associated with a drastic increasing of the uncertainty on position due to the 
space stretching\cite{N2012}\cite{Rodrigo}. In other words, we say that SSR allows to make the connection between the quantum uncertainties and 
the curvature of spacetime, thus leading to a geometrization of quantum mechanics to be deeply explored elsewhere. 

 On the other hand, the positive (attractive) gravity generates a positive pression for smashing the objects, thus leading to a decreasing 
of the uncertainty on position or a space contraction (positive curvatures), whereas the anti-gravity causes a negative pression in the
sense of the delocalization or an increase of the uncertainty on position\cite{N2015}\cite{N2012}. In sum, the barrier $c$ generates positive
curvatures (smashing) and so we have a more accurate localization of the particle, while the barrier $V$ generates negative curvatures 
(stretching) and thus we have an increase of the uncertainty on position. Such symmetry of SSR allows us to understand the origin of the 
uncertainty principle within the spacetime scenario\cite{N2012}.

 Inspired by Aldrovandi works\cite{Aldro1, Aldro2, Aldro3}, we calculate the Riemann tensor for the conformal metric in de Sitter 
Relativity, namely: 

\begin{equation}
 R^{\mu}_{\nu\rho\sigma}=\frac{\Lambda}{6}[\delta^{\mu}_{\rho}g_{\nu\sigma}-\delta^{\mu}_{\sigma}g_{\nu\rho}].
\end{equation}
 
 We should apply the result above to the extended dS-scenario\cite{Rodrigo}. To do that, we can calculate the Riemann tensor with
 functional dependence in terms of speed by representing the particle coupled to the background field 
 $\mathcal{G}_{\mu\nu}=\Theta(v)\eta_{\mu\nu}$. So we write 

\begin{equation}\label{ricci}
 \mathcal{R}^{\mu}_{\nu\rho\sigma}=\frac{\Lambda\Theta(v)}{6}[\delta^{\mu}_{\rho}\eta_{\nu\sigma}-\delta^{\mu}_{\sigma}\eta_{\nu\rho}], 
\end{equation}
where $\Theta(v)=1/(1-V^2/v^2)$, with $\Lambda(v)=\Lambda\Theta(v)$\cite{Rodrigo}. 

 By making $\sigma=\mu$ in Eq.(32), we find the Ricci tensor, namely: 

\begin{equation} 
 \mathcal{R_{\mu\nu}}(v)=-\frac{\Lambda\Theta(v)}{6}\eta_{\mu\nu}=-\frac{\Lambda}{6}\mathcal{G}_{\mu\nu}(v), 
\end{equation}
which represents the negative curvature generated by the cosmological constant. 

After obtaining the Ricci tensor in Eq.(33), we can use Einstein equation $R_{\mu\nu}-\frac{R}{2}g_{\mu\nu}=\Lambda_{\mu\nu}$ 
for calculating the energy-momentum tensor, where we have a energy-momentum tensor that is proportional to the metric. 
This is a tensor that characterizes the cosmological constant, since we will show that $\Lambda_{\mu\nu}\propto\Lambda g_{\mu\nu}$,
where $g_{\mu\nu}=\Theta\eta_{\mu\nu}=\mathcal G_{\mu\nu}$ in spacetime of SSR, which represents an extended dS-relativity\cite{Rodrigo}. 
 
 We know that the Ricci scalar is $R=2\Lambda$\cite{Rodrigo} in dS-relativity. By introducing such result and Eq.(33) into Einstein
equation, and also by considering $g_{\mu\nu}=\Theta\eta_{\mu\nu}=\mathcal G_{\mu\nu}$, so after performing the calculations, we find

\begin{equation}\label{tensor}
 \Lambda_{\mu\nu}=-\frac{7}{6}\Lambda\mathcal{G}_{\mu\nu},
\end{equation}
where we verify that the tensor $\Lambda_{\mu\nu}$ characterizes the cosmological constant $\Lambda$, and it is in fact proportional
to SSR-metric $\mathcal{G}_{\mu\nu}$, which is a solution of Einstein equation without sources and with the cosmological constant as shown
in a previous work\cite{Rodrigo}. 

Due to the presence of Lorentz symmetry breaking, we know that the covariant tensors are different from the contravariant ones. Here 
the tensors $\Lambda_{\mu\nu}$ and $\Lambda^{\mu\nu}$ provide different physical situations, however both of them respect the conformal 
symmetry. This has a connection with the emergent gravity theory\cite{pad,pad2}. In the case of SSR that breaks down Lorentz symmetry
close to the invariant minimum speed ($v\rightarrow V$) connected to the vacuum potential $\phi\rightarrow -c^2$, the covariant and 
contravariant metrics present opposite behaviors in such a limit, namely:   

\begin{equation}\label{1}
 \mathcal{G}_{\mu\nu}\rightarrow\infty
\end{equation}

and, on the other hand, in this limit $\phi\rightarrow -c^2$, the contravariant metric is 

\begin{equation}\label{2}
 \mathcal{G}^{\mu\nu}\rightarrow 0,  
\end{equation}
where we have $\mathcal{G}_{\mu\nu}=\Theta(v)\eta_{\mu\nu}$ [Eq.(29)] and $\mathcal{G}^{\mu\nu}=\Theta(v)^{-1}\eta^{\mu\nu}$,
being $\eta_{\mu\nu}=\eta^{\mu\nu}$. 

This situation occurs also in de Sitter spaces\cite{Aldro2}. Eq.(35) represents an infinite stretching of spacetime, which is  
associated with a complete delocalization of the particle, leading to a quantum uncertainty on position whose origin is the 
anti-gravity effect in spacetime of SSR ($\phi<0$). This kind of quantum behavior in the structure of spacetime
could symbolize an entanglement that is very similar to the phenomenon predicted in the horizons of black holes\cite{cool}. This 
issue deserves a deeper investigation elsewhere.

We already know that the curvature of SSR becomes very small (almost zero) in the approximation $v\approx v_0$, where we have a 
quasi-minkowskian space (see Fig.4). In order to perceive such an approximation, let us expand the metric of SSR in power series 
of $V/v$, namely: 

\begin{equation}\label{ex}
\mathcal{G}_{\mu\nu}=\eta_{\mu\nu}+\frac{V^2}{v^2}\eta_{\mu\nu}+\frac{V^4}{2!v^4}\eta_{\mu\nu}+\frac{V^6}{3!v^6}\eta_{\mu\nu}
+\frac{V^8}{4!v^8}\eta_{\mu\nu}+...
\end{equation}
Now we perceive that, for $v\approx v_0(>>V)$, we can neglect all the term of power of $V/v$ and so we simply find 
$\mathcal{G}_{\mu\nu}=\eta_{\mu\nu}$, which corresponds to the Minkowski space.

When truncating the series on the third term, we get $\mathcal{G}_{\mu\nu}=\eta_{\mu\nu}+
\frac{V^2}{v^2}\eta_{\mu\nu}+\frac{V^4}{2!v^4}\eta_{\mu\nu}$. 

If we make the approximation $\phi(v\rightarrow v^{-}_0)\approx 0$ [Eq.(17)], we find the covariant metric, as follows:  

\begin{equation}
\mathcal{G}_{\mu\nu}\approx\eta_{\mu\nu}, 
\end{equation} 
and the contravariant metric in this region is 

\begin{equation}
\mathcal{G}^{\mu\nu}\approx\eta^{\mu\nu}.
\end{equation}

We realize the both covariant and contravariant metrics are practically equal in this region of phase transition, so that Lorentz 
symmetry is restored in such a regime.
 
Alternatively, we can also adopt an expansion of the metric in terms of potential $\phi$ in such a way that in the vicinity of the 
phase transition, we find 

\begin{equation} 
\mathcal{G}_{\mu\nu}\approx\eta_{\mu\nu}-2\frac{\phi}{c^2}\eta_{\mu\nu},
\end{equation}
which is the covariant form of the metric in the first approximation close to the phase transition, where $\phi<0$, being
too close to zero, i.e., $\phi\approx 0^{-}$.   

And we also find  

\begin{equation}
\mathcal{G}^{\mu\nu}\approx\eta^{\mu\nu}+2\frac{\phi}{c^2}\eta^{\mu\nu}, 
\end{equation}
which is the contravariant form of the metric in the first approximation.

We see that, in this approximation, Lorentz symmetry is weakly broken, however the validity of Lorentz symmetry is still preserved
in such a first approximation, since $2\phi/c^2\approx 0$.  

Rewriting Eq.(40) and Eq.(41) in terms of knematics structure and making $v\approx v_0=\sqrt{cV}$, we find the covariant metric 
as follows: 

\begin{equation}
 \mathcal{G}_{\mu\nu}\approx\eta_{\mu\nu}+\frac{V}{c}\eta_{\mu\nu},
\end{equation}
and the contravariant metric, namely:

\begin{equation} 
 \mathcal{G}^{\mu\nu}\approx\eta^{\mu\nu}-\frac{V}{c}\eta^{\mu\nu}. 
\end{equation}

The dimensionless quantity $V/c=\xi$ is the gravito-electromagnetic coupling constant\cite{N2016}. 

We should stress that the EOS for vacuum energy ($p=-\rho$) obtained by SSR\cite{N2016} indicates that we are dealing with strongly
inflationary fluids when $\phi\approx -c^2$. 

 We know that SSR presents a dispersion relation\cite{N2016, N2015, N2012}, namely: 
  
\begin{equation}
 \mathcal P^{\mu}\mathcal P_{\mu}=m_0^2c^2\left(1-\frac{V^2}{v^2}\right). 
\end{equation} 

Multiplying the both sides of Eq.(44) by $\eta^{\mu\nu}$, we obtain

\begin{equation}
 \mathcal P^{\mu}\mathcal P^{\nu}=m_0^2c^2\left(1-\frac{V^2}{v^2}\right)\eta^{\mu\nu}.
\end{equation}

We can rewrite Eq.(45), as follows: 

 \begin{equation}
 \mathcal{P}^{\mu}\mathcal{P}^{\nu}=m_0^2c^2\mathcal{G}^{\mu\nu}, 
 \end{equation}
where $\mathcal{G}^{\mu\nu}=\Theta(v)^{-1}\eta^{\mu\nu}=(1-V^2/v^2)\eta^{\mu\nu}$.

Eq.(46) is very similar to the conformal relation of de Sitter\cite{banerj}. From Eq.(46), we realize that the momentum 
$\mathcal{P}^{\mu}$ in SSR obeys the following conformal relation: 

\begin{equation}
\mathcal{P}^{\mu}=\sqrt{1-\frac{V^2}{v^2}}p^{\mu}, 
\end{equation} 
with $p^{\mu}=m_{0}cU^{\mu}$, where $U^{\mu}=[\gamma,\frac{v_{i}}{c}\gamma]$ is the $4$-velocity of SR, but, in the approximation
$v<<c$, we can write $U^{\mu}\approx [1,\frac{v_{i}}{c}]$ with $i=1,2,3$. 

\subsection{The concept of reciprocal velocity and its connection with the uncertainty in position: the non-locality due to an apparent
tachyonic signal}

As already discussed in a previous paper\cite{N2012}, SSR generates a kinematics of non-locality as also proposed in the emergent
gravity theories\cite{acus2}. 

In order to see more clearly the aspect of non-locality of SSR due to the stretching of space when $v\approx V$, we should take into
account the idea of the so-called reciprocal velocity $v_{rec}$, which has been already well explored in a previous paper\cite{N2012}. 
Thus, here we will just reintroduce such idea in a more summarized way. To do that, let us use Eq.(3) and first multiply it by $c$
at both sides, and after by taking the squared result in order to obtain   

\begin{equation}
 \left(c^2-\frac{c^2V^2}{v^2}\right)(\Delta\tau)^2=(c^2-v^2)(\Delta t)^2, 
\end{equation}
where the right side of Eq.(48) related to the improper time $\Delta t$ provides the velocity $v$ of particle 
($\Delta t\rightarrow\infty$ for $v\rightarrow c$) and, on the other hand, the left side gives us the new information that shows that 
the proper time in SSR is not an invariant quantity as is in SR, so that the proper time goes to infinite ($\Delta\tau\rightarrow\infty$)
when $v\rightarrow V$, which leads to a too large stretching of the proper space interval $c\Delta\tau$ in this limit of much lower speed,
giving us the impression that the particle is well delocalized due its ``high internal speed" that is so-called reciprocal velocity that 
appears at the left side of the equation as being $v_{rec}=(cV)/v=v_0^2/v$. So we have $(c^2-v_{rec}^2)(\Delta\tau)^2=
(c^2-v^2)(\Delta t)^2$. Now we can perceive that the reciprocal velocity $v_{rec}$ represents a kind of inverse of $v$ such that, 
when $v\rightarrow V$, we get $v_{rec}\rightarrow c$, i.e., the ``internal motion'' is close to $c$, thus leading to the new effect 
of {\it proper time dilation} associated to a delocalization that was shown as being an uncertainty on position\cite{N2012} within
the scenario of spacetime in SSR. In this scenario, it was shown that the decreasing of momentum close to zero ($v\approx V$) 
leads to a delocalization of the particle, which is justified by the increasing of $v_{rec}\rightarrow c$ and the dilation of the proper 
time $\Delta\tau\rightarrow\infty$. As the uncertainty on position is $\Delta x=v_{rec}\Delta\tau=(v_0^2/v)\Delta\tau$\cite{N2012}, 
for $v\rightarrow V$, we find $\Delta x=c\Delta\tau\rightarrow\infty$. And, on the other hand, the large increasing of momentum for $v\rightarrow c$ 
leads to the well-known dilation of the improper time $\Delta t$ (right side of Eq.(48)), so that we find $\Delta\tau<<\Delta t$ and
the minimum reciprocal velocity $v_{rec}\rightarrow v_0^2/c=V$, which provides a small uncertainty on position, since 
$\Delta x=V\Delta\tau$.

Therefore Eq.(48) or Eq.(3) can be rewritten in the following way: 

\begin{equation}
\Delta\tau\sqrt{1-\frac{v^2_{rec}}{c^{2}}}=\Delta t\sqrt{1-\frac{v^{2}}{c^{2}}},
\end{equation}
where we find $v^2_{rec}/c^2=V^2/v^2$. We have $V<v<c$ and $V<v_{rec}<c$, where $V$ is the reciprocal of $c$ and vice versa.  
 
We claim that the concept of reciprocal velocity\cite{N2012} emerges in order to understand the apparent superluminal effects 
of tachyons as being the result of a new causal structure of spacetime with the presence of
a dark cone (dark energy) by representing $V$ (Fig.1), so that the so-called tachyonic inflation is now explained as being an effect of
a drastic proper time dilation and large stretching of the space close to the vacuum regime governed by the phase of anti-gravity (Fig.4), 
leading to an inflation (section 5). Thus, according to the idea of reciprocal velocity, when we are close to $V$ ($S_V$), we find 
$v_{rec}(v\approx V)\Delta\tau=(v_0^2/V)(\Delta\tau)=(v_0^2/V)(\Delta t)\theta^{-1}(v\approx V)=[c\theta^{-1}(v\approx V)]\Delta t=
c^{\prime}\Delta t\rightarrow\infty$, giving us the impression of a superluminal speed $c^{\prime}=c\theta^{-1}>>c$, since we now
consider that the dilation factor $\theta^{-1}=(1-V^2/v^2)^{-1/2}$ is acting on $c$ instead of acting on $\Delta t$, but the result
$c^{\prime}=c\theta^{-1}$ is merely apparent, working like a tachyon. Actually SSR shows that occurs a stretching of space.  

In other words, in sum we say that the tachyonic inflation ($v=c^{\prime}>>c$) just mimics the new spacetime effects of SSR close to 
the invariant minimum speed $V$. A deeper investigation of this issue will be made in the next section.

\section{Tachyonic fluids within the machian scenario of SSR and the breaking of the conformal property in the inflationary
scenario}

 It is already known that the relativistic Lagrangian of a free particle is 

\begin{equation}
 \mathcal{L}=-m_{0}c^{2}\sqrt{1-\frac{v^2}{c^2}}, 
\end{equation}
but, if the particle suffers the effects of a conservative force, which is independent of speed, we have 
$\mathcal{L}=-m_{0}c^{2}\sqrt{1-\beta^{2}}-U$, where $U=U(r)$ is a potential depending on position. If $L$ is a function that does not 
depend on time, so there is a constant of motion $h$, namely: 

\begin{equation}
 h=\dot q_{j} p_{j}-\mathcal{L}=\frac{m_0v_{j}v_{j}}{\sqrt{1-\frac{v^2}{c^2}}}+m_{0}c^{2}\sqrt{1-\frac{v^2}{c^2}}+U,  
\end{equation}
which leads to 
\begin{equation}
 h=\frac{m_{0}c^2}{\sqrt{1-\frac{v^2}{c^2}}}+U=E, 
\end{equation}
where $h$ is the total energy. If $U=0$, we have $E=\gamma m_0c^2$, which is the energy (constant of motion) of a free 
particle. We simply write $v^2=v_jv_j$ for a single particle. 

Now, by using an analogous procedure to that for obtaining the relativistic Lagrangian, we can find the Lagrangian of SSR. 
Let us first obtain the Lagrangian of a free particle. So, by considering $h=E=\Psi m_0c^2$ as a constant of motion for
a free particle in SSR and knowing that $p=\Psi m_0v$ is the momentum in SSR, we write 

\begin{equation}
 E=m_{0}c^{2}\sqrt{\frac{1-\frac{V^2}{v^2}}{1-\frac{v^2}{c^2}}}=h=m_{0}v^{2}\sqrt{\frac{1-\frac{V^2}{v^2}}{1-\frac{v^2}{c^2}}}
-\mathcal{L}, 
\end{equation}
where $v^2=v_jv_j$ and $\mathcal{L}$\cite{N2016} is the Lagrangian of a free particle in SSR.

From Eq.(53), we extract 

\begin{equation}
 \mathcal{L}=-m_{0}c^{2}\theta\sqrt{1-\frac{v^2}{c^2}}=-m_{0}c^{2}\sqrt{\left(1-\frac{V^2}{v^2}\right)\left(1-\frac{v^2}{c^2}\right)}.
\end{equation} 
If we make $V\rightarrow 0$ (or $v>>V$) in Eq.(54), we recover the relativistic Lagrangian of a free particle in Eq.(50). 

In the presence of a potential $U=U(r)$, we write

\begin{equation}
\mathcal{L}=-m_{0}c^{2}\sqrt{\left(1-\frac{V^2}{v^2}\right)\left(1-\frac{v^2}{c^2}\right)}-U. 
\end{equation}

Now let us consider the case of a particle in the presence of an electromagnetic field. In this case\cite{N2016}, 
we have a non-conservative potential that depends on position and speed. So we write

\begin{equation}
\mathcal{L}=-m_{0}c^{2}\sqrt{\left(1-\frac{V^2}{v^2}\right)\left(1-\frac{v^2}{c^2}\right)}-q\Phi+\frac{q}{c}\vec{A}\cdot\vec{v}. 
\end{equation}

If the charge $q=0$, we recover Eq.(54) that represents the Lagrangian for a free particle in SSR. 

We intend to show that the Lagrangian in Eq.(54) generates Machian effects\cite{mach, mach2} when the speed $v$ is too close to
the minimum speed $V$ connected to the ultra-referential $S_V$, thus leading to a strong coupling of the particle with vacuum, so that 
its mass becomes strongly dressed by the own vacuum energy. Thus, as the particle approaches more and more to the vacuum regime $S_V$, 
its coupling with vacuum increases drastically and so, its dressed mass increases a lot, such that the particle becomes extremely 
heavy due to the global effect of vacuum energy around it. Such effect so-called dressed mass\cite{N2015} has a Machian origin
within a quantum scenario of graviting vacuum. The dressed mass $m_{dress}$ of a particle shown in the approximation of vacuum regime 
given by the mimimum speed $V$ was investigated before in the ref.\cite{N2015}, where we have found 
$m_{dress}\approx m_0/\sqrt{1-V^2/v^2}$. On the other hand, SSR predicts that the relativistic (bare) mass $m\approx m_0\sqrt{1-V^2/v^2}$
goes to zero when $v\rightarrow V$, so that the ``particle'' loses its identity close to the vacuum-$S_V$ ($m\approx 0$) and so it becomes
completely delocalized (uncertainty on position as $c\Delta\tau\rightarrow\infty$)\cite{N2012} since it merges with vacuum as a whole,
leading to $m_{dress}\rightarrow\infty$. Actually, the impossibility of reaching the mimimum speed is due to the increasing of 
the dressed mass that goes to infinite for $v\rightarrow V$ and thus prevents the increasing of its decelaration in such a way that 
a too delocalized and so heavy particle cannot reach the minimum speed $V$. 

It is important to notice that the increasing of the dressed mass $m_{dress}$ at lower energies when $v\approx V$, due to the effect of
the vacuum field that couples strongly to the mass of the particle has a similarity to what occurs with QCD at lower energies, since, 
in such a confined regime, the constituent (bare) mass of a quark becomes dressed by the gluons, thus leading to an increasing of 
its mass. In this sense, we realize that the graviting vacuum within a Machian scenario has a non-local aspect, so that the
particles become confined by the own cosmological vacuum, which is directly connected to the metric of SSR by means of its conformal 
factor $\Theta(v)$\cite{N2016}\cite{Rodrigo} given when $v\approx V$. Therefore, thanks to the quantum Machian scenario of SSR that
leads to the emergence of the dressed mass, the whole universe works like a too huge ``bag'' where matter is confined by vacuum, 
as well as also occurs with the dressed quarks confined inside a proton. Such interesting similarity between SSR-vacuum and QCD-vacuum 
at lower energies, by dressing the masses of the particles deserves to be deeply investigated elsewhere.    
   
\subsection{The Dirac-Born-Infeld Lagrangian for a tachyonic fluid and its connection with SSR-Lagrangian within a Machian scenario}
 
The Dirac-Born-Infeld (D-B-I) Lagrangian\cite{li},\cite{rola, rola2, rola3, dark} that presents identical characteristics of a tachyonic fluid 
is of the type $L=\mathcal{V}(\phi)\sqrt{1-(\partial_{t}\phi)^2}$, where $\phi$ is a scalar field and $\mathcal V(\phi)$ is a tachyonic
potential. The potential $\mathcal{V}(\phi)$ is known as the Dirac-Born-Infeld potential. We should perceive that such a Lagrangian is 
similar to SSR-Lagrangian of a free particle given in Eq.(54), since we have already considered that the speed $v$ plays the role of a 
scalar field related to the gravitational potential $\phi$ of SSR [Eq.(15) and Eq.(16)]. 

Here we intend to go deeper in exploring the connection between D-B-I Lagrangian and SSR-Lagrangian by reinterpreting the tachyonic 
inflation (superluminal effects) within a new Machian scenario of graviting vacuum generated by spacetime of SSR when the concept 
of dressed mass naturally emerges from the scenario of D-B-I when compared with the scenario of SSR. To do that, by comparing 
D-B-I Lagrangian with SSR Lagrangian given in Eq.(54), we can identify the D-B-I potential $\mathcal{V}$ as being the term of SSR 
Lagrangian that provides the negative potential $-c^2$ multiplied by the factor $\theta(v)$, namely: 

\begin{equation}
 \mathcal{V}=-c^2\theta=-c^{2}\sqrt{1-\frac{V^2}{v^2}}\equiv-c^{2}\left(1+\frac{\phi}{c^2}\right),  
\end{equation}
where we have $v\approx V$ for $\phi\approx -c^2$, $\mathcal{V}$ being the D-B-I potential within SSR-scenario. 

The reason of considering $\theta$ for obtaining $\mathcal{V}$ is due to the fact that just $\theta$ allows to get corrections close
to vacuum ($v\approx V$), since vacuum is responsible for an inflation governed by anti-gravity.

As $\mathcal{V}$ in Eq.(57) is a kind of D-B-I potential within SSR-scenario, we can rewrite SSR Lagrangian in Eq.(54) in the 
following way: 

\begin{equation}
 \mathcal{L}=m_{0}\mathcal V\sqrt{1-\frac{v^2}{c^2}}, 
\end{equation}
where $\mathcal V=\mathcal V(v)\equiv\mathcal V(\phi)$ according to Eq.(57). As we already know that the speed $v$ also plays the
role of a scalar field, let us first consider the D-B-I potential as a function $\mathcal V=\mathcal V(v)$ for our proposal in 
finding the Machian scenario that should emerge from Eq.(57). And, since such Machian scenario is within the context of graviting 
vaccum for $v\approx V$, let us neglect the Lorentz factor in Eq.(58) by making the following approximation:  

\begin{equation}
\mathcal{L}\approx m_{0}\mathcal V(v), 
\end{equation}
where $\mathcal V(v)=-c^2\theta(v)$. 

Now, let us obtain the equation of motion from the Lagrangian given in Eq.(59). Therefore, we should use Euler-Lagrange equation. 
As $\mathcal L$ just has dependence on speed $v$, we simply find   

\begin{equation}
\frac{d}{dt}\left(\frac{\partial\mathcal L}{\partial v}\right)= m_0\frac{d}{dt}\left(\frac{\partial\mathcal V(v)}{\partial v}\right)=0, 
\end{equation}
from where we get the generalized momentum $\mathcal P$ of SSR given at lower energies, namely: 

\begin{equation}
\mathcal P= m_0\frac{\partial\mathcal V}{\partial v}=-m_0c^2\frac{d\theta(v)}{dv}, 
\end{equation}
where $\theta(v)=\sqrt{1-V^2/v^2}$. So, by introducing $\theta(v)$ into Eq.(61), we find  

\begin{equation}
\mathcal P=\mathcal P_{machian}=\frac{m_0}{\sqrt{1-\frac{V^2}{v^2}}}\left(\frac{c^2V^2}{v^3}\right),  
\end{equation}
which can be written in the following way: 

\begin{equation}
\mathcal P_{machian}=m_{dress}\left(\frac{v_0^4}{v^3}\right),  
\end{equation}
where $v_0=\sqrt{cV}$ and we find that the generalized momentum $\mathcal P$ depends directly on the dressed mass 
$m_{dress}=m_0/\sqrt{1-V^2/v^2}$\cite{N2015} as expected from D-B-I theory within SSR scenario. Therefore, we conclude that 
the generalized momentum $\mathcal P_{machian}$ is interpreted as being of Machian origin within a scenario of graviting vacuum, since
such Machian momentum increases drastically to infinite when $v\rightarrow V$ due to a very strong coupling of the particle $m_0$ with 
graviting vacuum at the universal reference frame $S_V$, thus leading to an effective mass dressed by vacuum, $m_{dress}$ being 
much larger than the bare mass $m_0$ given for $v=v_0$, i.e., we get $m_{dress}(v\approx V)>>m_0(v=v_0)$. So, we 
can finally conclude that a particle too close to the vacuum $S_V$ is extremely heavy and carries out a very high generalized
momentum that can be obtained from Eq.(63) by making $v\approx V$, so that we can write 

\begin{equation}
\mathcal P_{(v\approx V)}\approx m_{dress}\left(\frac{v_0^4}{V^3}\right),  
\end{equation}

or we can alternatively write this high generalized momentum of a ultra-heavy virtual particle of vacuum, as follows: 

\begin{equation}
\mathcal P_{vac}=m_{dress}(\sigma c),  
\end{equation}
where we have $c^{\prime}=\sigma c=(c/V)c=\xi^{-1}c\sim 10^{30}m/s$, with $\sigma=\xi^{-1}=c/V\sim 10^{22}$ and $c\sim 10^{8}m/s$. 

According to Eq.(65), we realize that a particle that is too close to the vacuum $S_V$ merges with the own vacuum fluid having a dressed 
mass that ``travels" as if it were a tachyon moving with the highest superluminal speed $c^{\prime}=\sigma c\sim 10^{30}m/s$ associated 
with the idea of ``omnipresence" of the ultra-referential $S_V$, so that, if we admit the Hubble radius $r_H(\sim 10^{26}m)$ for the
universe, such a tachyonic fluid with speed $\sigma c(\sim 10^{30}m/s)$ [Eq.(65)] would propagate across the whole universe during only 
about $r_H/\sigma c\sim 10^{-4}s$! However, in view of SSR scenario, it is important to stress that the superluminal effects of the
tachyonic fluids are apparent in the sense that such effects are in fact generated by a large stretching of spacetime when $v$ is too
close to $V$ (vacuum regime), so that we find $\Delta x^{\prime}=c(\Delta\tau)\approx c(\Delta t/\sqrt{1-V^2/v^2})\rightarrow\infty$
(see Eq.(3) for $v\approx V$), giving us an impression that we have a tachyonic inflation associated with a superluminal signal 
$c^{\prime}=c/\sqrt{1-V^2/v^2}\equiv c/(1+\phi/c^2)$, where we have a repulsive potential $-c^2<\phi<0$.

Actually we have realized that the D-B-I potential $\mathcal V$ generates a tachyonic fluid of vacuum with a superluminal speed $\sigma c$ 
that just mimics the spacetime of SSR where the large dimensionless factor $\sigma$ works like a scale factor for representing the space 
stretch, so that the speed of light is preserved. Such a stretching of space scale in SSR gives us the impression of the existence of
superluminal effects, but there are no tachyons in the SSR scenario. So the apparent superluminal speed $\sigma c$ is now responsible for 
the quantum Machian effects due to the non-locality and isotropy of the inflationary vacuum where the dimensionless constant $\sigma$ plays the role of
a scale factor that dilates drastically the space in vacuum regime.  

 Eq.(65) allows us to obtain the vacuum energy associated with its non-locality due to the large dimensionless scale 
factor $\sigma$, namely: 
    
\begin{equation}
 E_{vac}=\mathcal P(\sigma c)=m_{dress}\sigma^2 c^2.   
\end{equation}

 According to Einstein-Planck relation, we could associate the vacuum energy in Eq.(66) with a certain big quantum of energy $\hbar W$
due to the scale factor $\sigma$ that generates an amplified quantum of energy $\hbar W$. So we can write the following equivalence:  

\begin{equation}
 E_{vac}=m_{dress}\sigma^2 c^2=\hbar W, 
\end{equation}
where $W$ represents a very high frequency having origin in the non-local aspect of the vacuum $S_V$. 

Now, from Eq.(67), if we want to get the local aspect of a quantum of such vacuum energy (dark energy) without the effect of the 
scale factor $\sigma$, we just rewrite Eq.(67), as follows:  

\begin{equation}
 \epsilon_{vac}=m_{dress}c^2=\sigma^{-2}\hbar W=\xi^2\hbar W, 
\end{equation}

or simply 

\begin{equation}
 E_{darkion}=\xi^2 h\nu.
\end{equation}

Thus the momentum carried by this quantum of dark energy is 

\begin{equation}
 p_{darkion}=\frac{\xi^2 h\nu}{c}, 
\end{equation}
where $\xi^2=V^2/c^2\sim 10^{-44}$. The frequency is $\nu=W/2\pi$. $E_{darkion}$ represents the energy of a local quantum of vacuum (dark) energy 
with repulsive aspect, thus working like a kind of anti-graviton, but this conjecture will be 
deeper explored in the future. As this tiny quantum $\xi^2 h\nu$ comes from vacuum, let us denominate it as a {\it darkion}. Due to the so
small factor $\xi^2/c(\sim 10^{-52}m^{-1}s)$ in Eq.(70), such quantum (darkion) that has very low momentum $\xi^2 h\nu/c$ interacts very 
weakly with matter when compared with an electromagnetic field interacting with matter by means of a photon with momentum $h\nu/c$, since
we have a very low energy of a darkion $\xi^2 h\nu<<h\nu$ (photon). This is the reason why the minimal effects of such {\it darkion} (dark energy)
become so difficult to be detected locally, although the effects at cosmological scales for the accelerated expansion of the universe 
become measurable, since the factor $\sigma$ amplifies drastically the interaction with vacuum as a whole (Eq.66), i.e., 
the cosmological vacuum. This issue will be deeply explored elsewhere.  

\subsection{Lagrangian of SSR for a Machian scenario: the idea of dressed Lagrangian}

In view of the concept of reciprocal velocity $v_{rec}$\cite{N2012} reintroduced in the section 4, we can already realize that the 
Machian idea of dressed mass given in the scenario of SSR for vacuum regime ($v\approx V$), i.e., $m_{dress}=m_0/\sqrt{1-V^2/v^2}$
plays the role of a kind of reciprocal mass for the bare (relativistic) mass given in vacuum regime ($v\approx V$), i.e., 
$m\approx m_0\sqrt{1-V^2/v^2}$, since the bare mass is $m\rightarrow 0$ in the limit of $v\rightarrow V$ whereas the dressed (reciprocal)
mass $m_{dress}$ works like the inverse of $m$, i.e., $m_{dress}\rightarrow\infty$ for $v\rightarrow V$, which means an enormous effective 
mass as the result of the strong coupling of the particle with vacuum $S_V$ within this modern Machian scenario so-called quantum Machian
scenario due to the influence of the vacuum energy for creating masses as will be deeper explored soon when investigating the breaking
of the conformal property generated by the dressed Lagrangian of SSR in an inflationary scenario.
 
The existence of a dressed mass $m_{dress}$\cite{N2015} as the result of the interaction with the vacuum will allow us to obtain the
Lagrangian of interaction with the vacuum of SSR or the so-called dressed Lagrangian $\mathcal L_{dress}$ given close to
the vacuum regime. As we already know that the usual Lagrangian of a free particle close to $V$ is directly related to its bare mass, 
i.e., $\mathcal L=-(m_0\theta)c^2$, we will show that the dressed Lagrangian is $\mathcal L_{dress}=-(m_0\theta^{-1})c^2$, where 
$m_{dress}=m_0\theta^{-1}$. To do that, let us start from an analogous procedure as made before for obtaining $\mathcal L$ by using now the 
Legendre transformation for obtaining $\mathcal L_{dress}$, namely: 

\begin{equation}    
 h_{dress}=E_{dress}=p_{rec}\dot q-\mathcal L_{dress},
\end{equation}
where $\dot q=v$. 

 As we have $h_{dress}=E_{dress}=m_{dress}c^2=m_0c^2/\sqrt{1-V^2/v^2}$, $p_{rec}=m_{dress}v_{rec}=(m_0/\sqrt{1-V^2/v^2})(v_0^2/v)$ with
$v_{rec}=v_0^2/v=(cV)/v$, by substituting these results in Eq.(71), we find 

\begin{equation}
 E_{dress}=m_{dress}c^2=\frac{m_0c^2}{\sqrt{1-\frac{V^2}{v^2}}}=\frac{m_0v_0^2}{\sqrt{1-\frac{V^2}{v^2}}}-\mathcal L_{dress},
\end{equation}
where $v_0^2=cV$. 

From Eq.(72), we obtain 

\begin{equation}
 \mathcal L_{dress}= -\frac{m_0c^2(1-\xi)}{\sqrt{1-\frac{V^2}{v^2}}},
\end{equation}
where $\xi=V/c\sim 10^{-22}$. As this number is extremely small, we write $1-\xi\approx 1$. Thus let us neglect $\xi$ that will not 
affect our result and simply write 

\begin{equation}
\mathcal L_{dress}=-(m_0\theta^{-1})c^2=-m_{dress}c^2=-\frac{m_0c^2}{\sqrt{1-\frac{V^2}{v^2}}},
\end{equation}
which is exactly the result expected for $\mathcal L_{dress}$. 

\subsection{Breaking the conformal property in the inflationary scenario}

Let us first rewrite the Lagrangian in Eq.(59) in terms of the potential $\phi$, namely: 

\begin{equation}
\mathcal{L}=m_0\mathcal V(\phi)=-m_{0}c^{2}\Theta^{-\frac{1}{2}}(\phi)=-m_{0}c^{2}\left(1+\frac{\phi}{c^2}\right),
\end{equation}
where $\Theta^{-\frac{1}{2}}(\phi)=(1+\phi/c^2)$, with $-c^2<\phi<0$ by representing the repulsive potential.

By minimizing the Lagrangian in Eq.(75) via functional of Schwinger\cite{Carrol2}, we find the energy-momentum tensor of vacuum, 
as follows: 

\begin{equation}
 \mathcal{T}_{vac}^{\mu\nu}(\phi)=\frac{2}{\sqrt{-\mathcal{G}}}\frac{\delta(\sqrt{-\mathcal{G}}\mathcal{L})}
{\delta\mathcal{G}_{\mu\nu}}=2\Theta^{-2}(\phi)\frac{\delta(\Theta^{2}(\phi)\mathcal{L})}{\delta(\Theta(\phi))}\eta^{\mu\nu}, 
\end{equation}
where $\mathcal{G}_{\mu\nu}=\Theta(\phi)\eta_{\mu\nu}$, $\delta\mathcal{G}_{\mu\nu}=\delta(\Theta(\phi))\eta_{\mu\nu}$ and 
$\sqrt{-\mathcal{G}}=\Theta^{2}(\phi)$.

 By introducing Eq.(75) into Eq.(76), we obtain the contravariant energy-momentum tensor for a particle with speed $v\approx V$ or 
close to vacuum $S_V$, namely: 

\begin{equation}
 \mathcal{T}^{\mu\nu}_{vac}(\phi)=-3m_{0}c^{2}\Theta^{-\frac{3}{2}}(\phi)\eta^{\mu\nu}=
-3m_{0}c^{2}\left(1+\frac{\phi}{c^2}\right)\mathcal{G}^{\mu\nu}(\phi), 
\end{equation}
where $\mathcal{G}^{\mu\nu}(\phi)=\Theta(\phi)^{-1}\eta^{\mu\nu}=(1+\phi/c^2)^2\eta^{\mu\nu}$, such that we write Eq.(77) 
as follows: 

\begin{equation}
 \mathcal{T}^{\mu\nu}_{vac}(\phi)=-3m_{0}c^{2}\Theta^{-\frac{3}{2}}(\phi)\eta^{\mu\nu}=
-3m_{0}c^{2}\left(1+\frac{\phi}{c^2}\right)^3\eta^{\mu\nu},  
\end{equation}
where $\mathcal{T}^{\mu\nu}_{vac}$ represents the contravariant (bare) energy momentum-tensor for the vacuum in SSR. The dressed energy-momentum
tensor for vacuum will be obtained soon by means of the dressed Lagrangian given in Eq.(74). After this, we will interpret 
both results taking into account the meaning of both traces, namely the bare trace $\mathcal T$ and the dressed trace 
$\mathcal T_{dress}$ in the scenario of breaking of the conformal property.   

As we have defined $\eta^{\mu\nu}=\eta_{\mu\nu}=(1,-1,-1,-1)$ in Eq.(29), we obtain bare trace $\mathcal T$ of the contravariant tensor
$\mathcal{T}^{\mu\nu}_{vac}(\phi)$, namely:

\begin{equation}
 \mathcal T=6m_{0}c^{2}\left(1+\frac{\phi}{c^2}\right)^3,  
\end{equation}
from where we should have the repulsive phase of gravity $-c^2\leq\phi\leq 0$ (Fig.4) in order to obtain $0\leq \mathcal T\leq 6m_0c^2$.

Now we can rewrite the dressed Lagrangian in Eq.(74) in terms of the potential $\phi$, namely: 

\begin{equation}
\mathcal L_{dress}=-m_{0}c^{2}\Theta^{\frac{1}{2}}(\phi)=-m_{0}c^{2}\left(1+\frac{\phi}{c^2}\right)^{-1},
\end{equation}
where $\Theta^{\frac{1}{2}}(\phi)=(1+\phi/c^2)^{-1}$, with $-c^2<\phi<0$.

By minimizing the dressed Lagrangian in Eq.(80) via functional of Schwinger, we find the covariant energy-momentum tensor of vacuum,
namely:

\begin{equation}
 \mathcal{T}^{vac}_{\mu\nu}(\phi)=\frac{2}{\sqrt{-\mathcal{G}}}\frac{\delta(\sqrt{-\mathcal{G}}\mathcal{L})}
{\delta\mathcal{G}^{\mu\nu}}=2\Theta^{2}(\phi)\frac{\delta(\Theta^{-2}(\phi)\mathcal L_{dress})}{\delta(\Theta^{-1}(\phi))}\eta_{\mu\nu}, 
\end{equation}
where $\mathcal{G}^{\mu\nu}=\Theta^{-1}(\phi)\eta^{\mu\nu}$, $\delta\mathcal{G}^{\mu\nu}=\delta(\Theta^{-1}(\phi))\eta^{\mu\nu}$,  
$\sqrt{-\mathcal{G}}=\Theta^{-2}(\phi)$.

 Now by considering the change of variable $\Theta^{-1}=u$ with $\delta(\Theta^{-1})=\delta u$ and after by introducing Eq.(74) into 
Eq.(81), we obtain the covariant (dressed) energy-momentum tensor for a particle with speed $v\approx V$ or close to the vacuum $S_V$, 
namely:

\begin{equation}
 \mathcal{T}_{\mu\nu}^{vac}(\phi)=-3m_{0}c^{2}\Theta^{\frac{3}{2}}(\phi)\eta_{\mu\nu}=
-3m_{0}c^{2}\left(1+\frac{\phi}{c^2}\right)^{-1}\mathcal{G}_{\mu\nu}(\phi), 
\end{equation}
where $\mathcal{G}_{\mu\nu}(\phi)=\Theta(\phi)\eta_{\mu\nu}=(1+\phi/c^2)^{-2}\eta_{\mu\nu}$, such that we write Eq.(82) 
as follows: 

\begin{equation}
 \mathcal{T}_{\mu\nu}^{vac}(\phi)=-3m_{0}c^{2}\Theta^{\frac{3}{2}}(\phi)\eta_{\mu\nu}=
-3m_{0}c^{2}\left(1+\frac{\phi}{c^2}\right)^{-3}\eta_{\mu\nu},  
\end{equation}
where $\mathcal{T}_{\mu\nu}^{vac}$ represents the covariant (dressed) energy momentum-tensor for the vacuum in SSR.

As we have already defined $\eta^{\mu\nu}=\eta_{\mu\nu}=(1,-1,-1,-1)$, we obtain dressed trace $\mathcal T_{dress}$ of the covariant
tensor $\mathcal{T}_{\mu\nu}^{vac}(\phi)$, namely:

\begin{equation}
 \mathcal T_{dress}=\frac{6m_{0}c^{2}}{\left(1+\frac{\phi}{c^2}\right)^3}.  
\end{equation}

Eq.(79) and Eq.(84) provides the bare trace and dressed trace of the contravariant tensor $T_{vac}^{\mu\nu}$ and the covariant tensor 
$T^{vac}_{\mu\nu}$ respectively. It is important to note that both traces tend to zero and infinite respectively when we consider the most
repulsive potential $\phi\rightarrow -c^2$. So, in order to interpret both results within a Machian scenario of an inflationary universe 
with a certain horizon radius $r_u$, we should naturally connect such most repulsive potential with a positive cosmological parameter 
$\Lambda(r_u)$ that varies with $r_u$ for a fixed potential of vacuum too close to $-c^2$, so that we write $\phi\approx -c^2=
-\Lambda r_u^2/6$, from where we get $\Lambda(r_u)=6c^2/r_u^2>0$, allowing us to obtain the tiny value of the current cosmological
parameter that is known as being the cosmological constant $\Lambda_0=6c^2/r_H^2\sim 10^{-35}s^{-2}$ given for the vacuum energy 
($\phi=-c^2$) in the frontier of the sphere (universe) with Hubble radius $r_H\sim 10^{26}$m\cite{N2008}\cite{N2015}\cite{N2016}.

 As we are considering an eternal inflation\cite{Hawking} of a dark sphere (universe) that expands due to the vacuum potential $\phi\approx -c^2$, we always 
obtain $\mathcal T\approx 0$ and $\mathcal T_{dress}\rightarrow\infty$ at any horizon radius $r_u$, so that $\Lambda$ goes to zero
for $r_u\rightarrow\infty$. Thus, in the early universe, if we consider $r_u=L_P(\sim 10^{-35}m)$ (Planck radius), the cosmological
parameter was too huge given by $\Lambda_P=6c^2/L_P^2\sim 10^{88}s^{-2}$, which corresponds about $123$ orders of magnitude beyond
the current value of $\Lambda$ or the so-called cosmological constant. This led to the fastest inflationary period that occurred 
in the origin of our universe when the space increased very rapidly due to the extremely high energy density (pressure) of vacuum, i.e.,
$\rho_P=-p_P=\Lambda_P c^2/8\pi G\sim 10^{94}g/cm^3 (\equiv p_P=-10^{114}N/m^2)$, which corresponds about $123$ orders of magnitute beyond
the current vacuum density $\rho_0\sim 10^{-29}g/cm^3 (\equiv p_0=-10^{-9}N/m^2)$. 

  Let us write Eq.(79) and Eq.(84) in the following ways: 

 \begin{equation}
  \mathcal T^{-}=6m_{0}c^{2}\left(1-\frac{\Lambda r_u^2}{6c^2}\right)^3, 
 \end{equation}

and 

\begin{equation}
 \mathcal T^{+}=\frac{6m_{0}c^{2}}{\left(1-\frac{\Lambda r_u^2}{6c^2}\right)^3},    
\end{equation}
where $\phi=-\Lambda r_u^2/6$. 

Here we are fixing a certain vacuum potential $\phi$ close to $-c^2$, i.e., $\phi\approx -c^2$, so that we obtain an almost null value
for $\mathcal T^{-}$ ($T^{-}\approx 0$) and a very high value for $\mathcal T^{+}$. Thus, we can see the inflation of the 
dark sphere, beginning with a very high value of $\Lambda$, which decreases with time. 

At a first sight, this is a simple model of an inflationary sphere of dark energy. However, since Eq.(85) and Eq.(86) show us a breaking
of the conformal property due to the exponent $3$ of the factor $(1+\phi/c^2)$ instead of the exponent $2$ that appears in the 
conformal metric of SSR [Eq.(30)]\cite{Rodrigo}, both energy-momentum tensors for the vacuum are not conformal structures. 

The presence of an invariant minimum speed in spacetime also breaks the translational symmetry of the Poincar\'e group, since it was shown 
rigorously in two previous papers\cite{N2016}\cite{N2015} that the multiplicative neutral element given by the identity matrix in
the subgroup of Lorentz does not exist in the limit of vacuum ($v\approx V$), so that the violation of this fundamental condition
of neutral element related to the classical idea of rest (rotational invariance) also affects the translational invariance of the
Poincar\'e group, thus leading to a violation of the conservation of the vacuum energy-momentum tensor, which becomes more evident with 
the breaking of the conformal structure in Eq.(85) and Eq.(86). 

Therefore, now we can improve the interpretation for this simple model of an inflationary dark sphere (universe), by taking into account 
very big fluctuations of vacuum energy during the inflation due to the non-conserved energy-momentum tensor, so that matter can be
widely created and destroyed by the vacuum by means of the quantities $\mathcal T^{+}$ and $\mathcal T^{-}$ respectively. We can note that 
such quantities are related to the cosmological paremeter $\Lambda$ that generates the cosmic inflation. We also notice that the
quantities $\mathcal T^{-}$ and $\mathcal T^{+}$ destroy and create matter by means of the same factor $(1+\phi/c^2)^3$ given 
in the nunerator and denumerator of Eq.(85) and Eq.(86) respectively, so that the average of matter becomes zero in a very short 
interval of time in the early universe. However, for a much longer period of time, new breakings of symmetries beyond the conformal symmetry 
breaking have ocurred, such that the quantity of matter created by vacuum overcame the quatity of matter destroyed
by vacuum and also the quantity of matter created overcame the quantity of anti-matter by means of a CPT violation, which is
not understood yet. This issue about other breakings of symmetry beyond the conformal symmetry breaking should be deeply investigated 
in the future. 

\section{Some prospects}

AdS-spaces also could emerge from this spacetime in the sector of $v>v_0$ that corresponds to the positive potentials $\phi$. This will
generate a kind of AdS-metric and its interesting implications to be investigated elsewhere. 
               
As the vacuuum energy in SSR generates a super-fluid being associated with the cosmological constant\cite{N2016}, thus leading to a 
de Sitter scenario\cite{Rodrigo} governed by anti-gravity, the spacetime of SSR allows us to obtain a phase transition between gravity
and anti-gravity. In view of this, in the future we will intend to address the puzzles associated with the gravitational collapse, by
taking into account the phase transition from gravity to anti-gravity in the spacetime with an invariant minimum speed. Perhaps, 
we should suspect that the presence of the dark cone for representing $V$ (Fig.1) could affect the gravitational collapse that normally 
leads to a classical black hole, but, due to the presence of the minimum speed $V$, we will investigate whether there could emerge
a very strong anti-gravity for a certain very small radius slightly larger than the Schwarzschild radius in a region of phase transition
predicted by SSR during the collapse, so that the event horizon would be prevented and thus a new collapsed structure could emerge.
Such a new collapsed structure could not be a physical singularity in the classical sense, but something that just could remind gravastars
or any kind of black hole mimickers (quantum black holes) that could also emit many types of radiations, probably by also including the 
gamma bursts. This intriguing issue deverves an exhaustive investigation in future papers.

 {\noindent\bf Acknowledgements} \\
I am grateful to Prof. Jonas Durval Cremasco and Giuseppe Vicentini. Rodrigo Francisco do Santos is grateful to Prof. J. A. Helayel-Neto (CBPF) and Prof. S\'ergio Uhloa (UNB) for interesting discussions. I, the father of SSR theory, dedicate this work to the memory of the great scientist Albert Einstein and also recently to the memory of the great scientist Stephen Hawking who has investigated exhaustively the nature of the black holes, which are still a big puzzle in the scenario of quantum gravity.

\end{document}